\documentclass{article}
\usepackage{amssymb}
\usepackage{graphics} 
\begin{document}
\begin{center}
{\large \bf Discrete quasiperiodic sets with predefined 
covering cluster}\\[3mm]
{N. COTFAS}\\[2mm]
Faculty of Physics, University of Bucharest, E-mail: ncotfas@yahoo.com
\end{center}
Some of the most remarkable tilings and discrete quasiperiodic sets used in 
quasicrystal physics can be obtained by using strip projection 
method in a superspace of dimension four, five or six, and the 
projection of a unit hypercube as a window of selection.
We present some mathematical results which allow one to use this very 
elegant method in superspaces of dimension much higher, and to generate 
discrete quasiperiodic sets with a more complicated local structure 
by starting from the corresponding covering cluster. Hundreds of points
of these sets can be obtained in only a few minutes by using our computer programs.\\[5mm]
{\bf 1. Introduction}\\[5mm]
Quasicrystals are materials with perfect long-range order, 
but with no three-dimensional translational periodicity.
The discovery of these solids in the early 1980's and the challenge 
to describe their structure led to a great interest in discrete quasiperiodic 
sets and their coverings (\cite{kp} and references therein).

The diffraction image of a quasicrystal
often contains  a set of sharp Bragg peaks invariant under a finite
non-crystallographic group of symmetries $G$, called the symmetry group 
of quasicrystal (in reciprocal space). 
In the case of quasicrystals with no translational periodicity this group
is the icosahedral group $Y$ and in the case of quasicrystals 
periodic along one direction (two-dimensional quasicrystals) $G$ is one 
of the dihedral groups $D_8$ (octagonal quasicrystals), $D_{10}$ 
(decagonal quasicrystals) and $D_{12}$ (dodecagonal quasicrystals).
Real structure information obtained by high resolution transmission 
electron microscopy suggests us that a quasicrystal with symmetry group
$G$ can be regarded as a quasiperiodic packing of  
copies of a well-defined $G$-invariant cluster $\mathcal{C}$. 

In the literature on quasicrystals the term `cluster' has several meanings \cite{kp}.
In the present paper,  by $G$-{\it cluster}  we mean a finite union 
of orbits of a finite group $G$ in a fixed representation. A mathematical algorithm 
for generating quasiperiodic point sets by starting from $G$-clusters 
was proposed by author in collaboration with Jean-Louis Verger-Gaugry several
years ago \cite{jlvg}. It is based on strip projection method 
(\cite{kp} and references therein) and is a direct generalization of the algorithm 
used by Katz and Duneau
in \cite{md}. The model obtained in \cite{md} for the icosahedral quasicrystals starts
from the one-shell $Y$-cluster $\mathcal{C}$ formed by the vertices of a regular
icosahedron. The physical space is embedded into the superspace $\mathbb{R}^6$ such 
that the orthogonal projections on the physical space of the points
\[ (\pm 1,0,0,0,0,0),\, (0,\pm 1,0,0,0,0),\, ...,\, (0,0,0,0,\pm 1,0),\, 
(0,0,0,0,0,\pm 1) \]
are the vertices of a regular icosahedron.
 
In our direct generalization, we consider only $G$-clusters invariant under inversion.
If our starting $G$-cluster $\mathcal{C}$ has $2k$ points then we embed the physical
space into the superspace $\mathbb{R}^k$ in such a way that 
$\mathcal{C}$ is the orthogonal projection on the physical space of the subset
\[ \{ (\pm 1,0,0,...,0),\, (0,\pm 1,0,0,...,0),\, ...,\, (0,0,...,0,\pm 1,0),\, 
(0,0,...,0,\pm 1)\} \]
of $\mathbb{R}^k$ containing $2k$ points. One can remark that, 
in the case of a two-shell or three-shell cluster, the dimension of the involved superspace 
is rather high.

Our aim is to present some mathematical results which allow one to use our algorithm 
in the case of multi-shell clusters. We show that in the case of a two-dimensional 
(resp. three-dimensional) cluster we have to compute only determinants of order 
three (resp. four), independently of the dimension of the superspace we use.
This remark and a simple description of the window (which, generally, is a polyhedron 
with hundreds or thousands faces) have allowed us to obtain some very efficient computer 
programs for our algorithm \cite{cp}.
 
In the case of a three-shell $Y$-cluster formed
by the vertices of a regular icosahedron, a regular dodecahedron and an icosidodecahedron
we use a 31-dimensional superspace, the window is a polyhedron lying in a 28-dimensional 
subspace bounded by 31465 pairs of 
parallel faces,  but we obtain 400-500 points in less than
10 minutes \cite{cp2}. In the case of a two-shell $D_{10}$-cluster we use a 10-dimensional
superspace, the window is bounded by 120 pairs of parallel faces, 
and we obtain 700-800 points in only one minute \cite{cp1}.\\[5mm]
{\bf 2. $G$-clusters}\\[3mm]
Consider a finite group $G$ and a fixed $\mathbb{R}$-irreducible representation of $G$
in $\mathbb{R}^n$.
In the case of dihedral groups 
$D_{2m}\!=\!\langle  a, \, b\ |\  a^{2m}\!=\!b^2\!=\!(ab)^2\!=\!e \rangle $
we can use the two-dimensional representation defined by 
$a,\, b :\mathbb{R}^2\longrightarrow \mathbb{R}^2$
\begin{equation}\begin{array}{l}
 a(\alpha ,\beta )=
\left(\alpha \, \cos \frac{\pi }{m}-\beta \, \sin \frac{\pi }{m}, \
      \alpha \, \sin \frac{\pi }{m}+\beta \, \cos \frac{\pi }{m} \right)\\[1mm]
b(\alpha ,\beta )=(\alpha , -\beta ) 
\end{array} \end{equation}
and in the case of the icosahedral group 
$Y\!=\!\langle a,\, b\ |\ a^5\!=\!b^2\!=\!(ab)^3\!=\!e \rangle $ the three-dimensional
representation generated by the rotations
$a,\, b :\mathbb{R}^3\longrightarrow \mathbb{R}^3$
\begin{equation}\label{Y} \begin{array}{l}
a(\alpha ,\beta  ,\gamma )=
\left(\frac{\tau -1}{2}\alpha -\frac{\tau }{2}\beta  +\frac{1}{2}\gamma ,
\ \frac{\tau }{2}\alpha +\frac{1}{2}\beta  +\frac{\tau -1}{2}\gamma ,
\ -\frac{1}{2}\alpha +\frac{\tau -1}{2}\beta  
+\frac{\tau }{2}\gamma \right)\\[1mm]
b(\alpha ,\beta  ,\gamma )=(-\alpha ,-\beta  ,\gamma ).
\end{array} \end{equation}
where $\tau =(1+\sqrt{5})/2$. 
The set $G(\alpha _1,\alpha _2,...,\alpha _n)
=\{ \  g(\alpha _1,\alpha _2,...,\alpha _n)\ |\ g\in G\ \}$ is called
the {\it orbit} of $G$ generated by $(\alpha _1,\alpha _2,...,\alpha _n)$. 
If $ \alpha \in (0,\infty )$  then the orbit
\[ \begin{array}{ll}
Y(\alpha ,\alpha \tau ,0) & 
{\rm is\ formed\ by\ \ the\ vertices\ of\ a\ regular\ icosahedron, }\\
Y(\alpha ,\alpha ,\alpha ) &
{\rm is\ formed\ by\ the\ vertices\ of\ a\ regular\ dodecahedron, }\\
Y(\alpha ,0,0) &
{\rm is\ formed\ by\  the\ vertices\ of\ an\ icosidodecahedron .}
\end{array} \]
For each $(\alpha , \beta )\not=(0,0)$, the orbit $D_{2m}(\alpha ,\beta )$
is formed by the vertices of a regular polygon with $2m$ sides.
A $G$-{\it cluster} is a finite union of orbits of $G$. For example, 
$\mathcal{C}=D_{2m}(\alpha _1,\beta _1) \cup D_{2m}(\alpha _2,\beta _2)$ is two-shell 
$D_{2m}$-cluster, and 
\[ \mathcal{C}=
Y(\alpha ,\alpha \tau ,0)\cup Y(\beta , \beta ,\beta )\cup Y(\gamma , 0,0)\}\]
is a three-shell icosahedral cluster.\\[5mm]
{\bf 3. Discrete quasiperiodic sets defined by $G$-clusters}\\[3mm]
Let \ $\mathcal{C}=\{ v_1,\, v_2,\, ...,\, v_k,\, -v_1,\, -v_2,\, ...,\, -v_k \}$, where 
$v_j=(v_{1j},v_{2j},...,v_{nj})$, 
be a fixed $G$-cluster symmetric with respect to the origin.  
One can prove \cite{jlvg,nc} that the vectors 
$w_1$, $w_2$, ..., $w_n$, where
$w_i\!=\!(v_{i1},v_{i2},...,v_{ik})$, are orthogonal and have the same norm 
(which we denote by $\kappa $)
\begin{equation}
\langle w_i,w_j\rangle =v_{i1}\, v_{j1}+v_{i2}v_{j2}+...+v_{ik}v_{jk}
=\left\{ \begin{array}{ll}
\kappa ^2 & {\rm if \ \ } i=j\\
0 & {\rm if\ \ } i\not= j.
\end{array} \right. 
\end{equation}
We identify the space $\mathbb{R}^n$ containing $\mathcal{C}$ (`physical space')
with the subspace
\begin{equation}
{\bf E}=\{ \ \alpha _1w_1+\alpha _2w_2+...+\alpha _nw_n\ 
|\ \alpha _1,\, \alpha _2,\, ...,\, \alpha _n\in \mathbb{R}\ \}.
\end{equation}
of the {\it superspace} $\mathbb{R}^k$, and consider the orthogonal complement of
${\bf E}$ 
\begin{equation} 
{\bf E}^\perp =\{ \ x\in \mathbb{R}^k\ |\ 
\langle x,y\rangle =0\ {\rm for\ all}\ y\in {\bf E}\ \}. 
\end{equation}
The orthogonal projectors corresponding to ${\bf E}$ and ${\bf E}^\perp $ are 
\begin{equation} \begin{array}{ll}
\pi :\mathbb{R}^k\longrightarrow {\bf E} & 
\pi \, x= \left\langle x,\frac{w_1}{\kappa }\right\rangle \frac{w_1}{\kappa }+
          \left\langle x,\frac{w_2}{\kappa }\right\rangle \frac{w_2}{\kappa }+...+
          \left\langle x,\frac{w_n}{\kappa }\right\rangle \frac{w_n}{\kappa }\\[1mm]
\pi ^\perp :\mathbb{R}^k\longrightarrow {\bf E}^\perp &  \pi ^\perp x=x-\pi \, x. 
\end{array} \end{equation}
If we describe ${\bf E}$ by using the orthogonal basis 
$\{ \kappa ^{-2}w_1,\, \kappa ^{-2}w_2,\, ...,\, \kappa ^{-2}w_n\}$ 
then the orthogonal projector corresponding to ${\bf E}$ becomes
\begin{equation} \mathcal{P}: \mathbb{R}^k\longrightarrow \mathbb{R}^n\qquad 
\mathcal{P}x=(\langle x,w_1\rangle , \langle x,w_2\rangle,...,
\langle x,w_n\rangle  ). \end{equation}

The projection ${\bf W}=\pi ^\perp (\Omega )$ of the unit hypercube
\[ \Omega =
\left[-\frac{1}{2},\, \frac{1}{2}\right]^k=
\left\{ (x_1,x_2,...,x_k)\ \left|\ -\frac{1}{2}\leq x_i\leq \frac{1}{2}\
{\rm for\ all\ } i\in \{ 1,2,..., k\}\ \right. \right\} 
\]
is a polyhedron (called a {\it window}) in the $(k-n)$-subspace ${\bf E}^\perp $. 
Each $(k-n-1)$-dimensional face of ${\bf W}$ is the 
projection of a $(k-n-1)$-face of the unit hypercube $\Omega $.
Each $(k-n-1)$-face of $\Omega $ is parallel to $k-n-1$ vectors of the canonical
basis $\{ e_1,\, e_2,\, ...,\, e_k\}$ of $\mathbb{R}^k$, and orthogonal to $n+1$ of them.
For each $n+1$ distinct vectors $e_{i_1}$, $e_{i_2}$, ..., $e_{i_{n+1}}$ the number
of $(k-n-1)$-faces of $\Omega $ orthogonal to them is $2^{n+1}$, and the set
\[ \left\{ \ x=(x_1,x_2,...,x_k)\ \left| \ \begin{array}{lcl}
x_i\in \{ -1/2,\, 1/2\} & {\rm if}&   i\in \{ i_1,\, i_2,\,...,\, i_{n+1}\} \\
x_i=0 & {\rm if}& i\not\in \{ i_1,\, i_2,\, ...,\, i_{n+1}\}
\end{array} \right. \right\} \]
contains one and only one point from each of them.

There are $k!/[(n\!+\!1)!\, (k\!-\!n\!-\!1)!]$ sets of $2^{n+1}$ parallel 
$(k\!-\!n\!-\!1)$-faces of 
$\Omega $. In the case $n=2$ these sets can be labelled by using the elements of 
\[ \mathcal{I}=
\{ (i_1,i_2,i_3)\in \mathbb{Z}^3\ |\ 1\leq i_1\leq k-2,\ \ i_1+1\leq i_2\leq k-1,\ \ 
                                  i_2+1\leq i_3\leq k\ \} \]
and in the case $n=3$ the elements of 
\[ \mathcal{I}=
\left\{ (i_1,i_2,i_3,i_4)\in \mathbb{Z}^4\ \left|\ 
            \begin{array}{rl}
            1\leq i_1\leq k-3,\ \ \ & i_1+1\leq i_2\leq k-2,\\ 
            i_2+1\leq i_3\leq k-1,\ \ \  & i_3+1\leq i_4\leq k
            \end{array} \right.
\right\}. \]

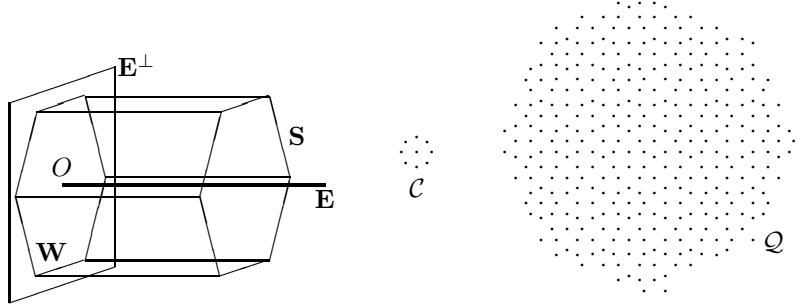
\begin{figure}
\setlength{\unitlength}{0.7mm}
\begin{picture}(130,60)(-5,0)
\put(62,18){${\bf E}$}
\put(80,20){$\mathcal{C}$}
\put(147,10){$\mathcal{Q}$}
\put(24.5,43){${\bf E}^\perp $}
\put(12,24){$O$}
\put(9,8){${\bf W}$}
\put(57,30){${\bf S}$}
\put(4,0){\line(0,1){38}}
\put(4,0){\line(3,1){20}}
\put(4,38){\line(3,1){20}}
\put(24,6.8){\line(0,1){38}}
\put(9,5){\line(3,1){9.1}}
\put(9,5){\line(1,0){35}}
\put(9,5){\line(-1,4){3.8}}
\put(5.2,20){\line(1,4){4}}
\put(5.2,20){\line(1,0){35}}
\put(18.4,8){\line(1,4){4}}
\put(18.4,8){\line(1,0){35}}
\put(22.3,23.9){\line(-1,4){3.8}}
\put(22.3,23.9){\line(1,0){35}}
\put(9.2,36.3){\line(3,1){9.1}}
\put(9.2,36.3){\line(1,0){35}}
\put(18.4,39.1){\line(1,0){35}}
\put(44,5){\line(3,1){9.1}}
\put(44,5){\line(-1,4){3.8}}
\put(40.2,20){\line(1,4){4}}
\put(53.4,8){\line(1,4){4}}
\put(57.3,23.9){\line(-1,4){3.8}}
\put(44.2,36.3){\line(3,1){9.1}}
\linethickness{0.4mm}
\put(14,22.3){\line(1,0){50}}
\setlength{\unitlength}{2mm}
\put(   44.30000,  10.72426){\circle*{0.2}} 
\put(   43.30000,  10.72426){\circle*{0.2}} 
\put(   43.59289,  10.01716){\circle*{0.2}} 
\put(   45.00711,  11.43137){\circle*{0.2}} 
\put(   44.30000,   9.72426){\circle*{0.2}} 
\put(   44.30000,  11.72426){\circle*{0.2}} 
\put(   45.00711,  10.01716){\circle*{0.2}} 
\put(   42.59289,  10.01716){\circle*{0.2}} 
\put(   43.30000,  11.72426){\circle*{0.2}} 
\put(   42.59289,  11.43137){\circle*{0.2}} 
\put(   43.59289,   9.01716){\circle*{0.2}} 
\put(   46.00711,  11.43137){\circle*{0.2}} 
\put(   45.00711,  12.43137){\circle*{0.2}} 
\put(   45.71421,  10.72426){\circle*{0.2}} 
\put(   45.00711,   9.01716){\circle*{0.2}} 
\put(   43.59289,  12.43137){\circle*{0.2}} 
\put(   46.00711,  10.01716){\circle*{0.2}} 
\put(   41.59289,  10.01716){\circle*{0.2}} 
\put(   42.59289,   9.01716){\circle*{0.2}} 
\put(   41.88579,  10.72426){\circle*{0.2}} 
\put(   42.59289,  12.43137){\circle*{0.2}} 
\put(   44.30000,   8.31005){\circle*{0.2}} 
\put(   46.00711,  12.43137){\circle*{0.2}} 
\put(   46.71421,  10.72426){\circle*{0.2}} 
\put(   45.71421,  13.13848){\circle*{0.2}} 
\put(   45.00711,  13.43137){\circle*{0.2}} 
\put(   44.30000,  13.13848){\circle*{0.2}} 
\put(   46.00711,   9.01716){\circle*{0.2}} 
\put(   45.71421,   8.31005){\circle*{0.2}} 
\put(   41.59289,   9.01716){\circle*{0.2}} 
\put(   40.88579,  10.72426){\circle*{0.2}} 
\put(   41.88579,   8.31005){\circle*{0.2}} 
\put(   43.30000,   8.31005){\circle*{0.2}} 
\put(   41.88579,  11.72426){\circle*{0.2}} 
\put(   41.59289,  12.43137){\circle*{0.2}} 
\put(   43.30000,  13.13848){\circle*{0.2}} 
\put(   42.59289,  13.43137){\circle*{0.2}} 
\put(   41.88579,  13.13848){\circle*{0.2}} 
\put(   44.30000,   7.31005){\circle*{0.2}} 
\put(   45.00711,   7.60294){\circle*{0.2}} 
\put(   46.71421,  13.13848){\circle*{0.2}} 
\put(   46.71421,  11.72426){\circle*{0.2}} 
\put(   47.71421,  10.72426){\circle*{0.2}} 
\put(   47.42132,  11.43137){\circle*{0.2}} 
\put(   46.71421,   9.72426){\circle*{0.2}} 
\put(   47.42132,  10.01716){\circle*{0.2}} 
\put(   45.71421,  14.13848){\circle*{0.2}} 
\put(   44.30000,  14.13848){\circle*{0.2}} 
\put(   46.71421,   8.31005){\circle*{0.2}} 
\put(   40.88579,   8.31005){\circle*{0.2}} 
\put(   40.88579,   9.72426){\circle*{0.2}} 
\put(   40.17868,  10.01716){\circle*{0.2}} 
\put(   40.88579,  11.72426){\circle*{0.2}} 
\put(   40.17868,  11.43137){\circle*{0.2}} 
\put(   41.88579,   7.31005){\circle*{0.2}} 
\put(   42.59289,   7.60294){\circle*{0.2}} 
\put(   43.30000,   7.31005){\circle*{0.2}} 
\put(   40.88579,  13.13848){\circle*{0.2}} 
\put(   43.30000,  14.13848){\circle*{0.2}} 
\put(   41.88579,  14.13848){\circle*{0.2}} 
\put(   43.59289,   6.60294){\circle*{0.2}} 
\put(   45.00711,   6.60294){\circle*{0.2}} 
\put(   46.00711,   7.60294){\circle*{0.2}} 
\put(   47.71421,  13.13848){\circle*{0.2}} 
\put(   46.71421,  14.13848){\circle*{0.2}} 
\put(   47.42132,  12.43137){\circle*{0.2}} 
\put(   48.42132,  11.43137){\circle*{0.2}} 
\put(   48.42132,  10.01716){\circle*{0.2}} 
\put(   47.42132,   9.01716){\circle*{0.2}} 
\put(   45.00711,  14.84558){\circle*{0.2}} 
\put(   43.59289,  14.84558){\circle*{0.2}} 
\put(   47.71421,   8.31005){\circle*{0.2}} 
\put(   46.71421,   7.31005){\circle*{0.2}} 
\put(   40.17868,   7.60294){\circle*{0.2}} 
\put(   40.88579,   7.31005){\circle*{0.2}} 
\put(   40.17868,   9.01716){\circle*{0.2}} 
\put(   39.17868,  10.01716){\circle*{0.2}} 
\put(   39.47157,  10.72426){\circle*{0.2}} 
\put(   40.17868,  12.43137){\circle*{0.2}} 
\put(   39.17868,  11.43137){\circle*{0.2}} 
\put(   42.59289,   6.60294){\circle*{0.2}} 
\put(   40.88579,  14.13848){\circle*{0.2}} 
\put(   42.59289,  14.84558){\circle*{0.2}} 
\put(   44.30000,   5.89584){\circle*{0.2}} 
\put(   46.00711,   6.60294){\circle*{0.2}} 
\put(   45.00711,   5.60294){\circle*{0.2}} 
\put(   47.71421,  14.13848){\circle*{0.2}} 
\put(   48.42132,  12.43137){\circle*{0.2}} 
\put(   47.42132,  14.84558){\circle*{0.2}} 
\put(   46.00711,  14.84558){\circle*{0.2}} 
\put(   49.12843,  10.72426){\circle*{0.2}} 
\put(   48.42132,   9.01716){\circle*{0.2}} 
\put(   45.00711,  15.84558){\circle*{0.2}} 
\put(   44.30000,  15.55269){\circle*{0.2}} 
\put(   47.71421,   7.31005){\circle*{0.2}} 
\put(   48.42132,   7.60294){\circle*{0.2}} 
\put(   47.42132,   6.60294){\circle*{0.2}} 
\put(   39.17868,   7.60294){\circle*{0.2}} 
\put(   40.17868,   6.60294){\circle*{0.2}} 
\put(   39.47157,   8.31005){\circle*{0.2}} 
\put(   41.59289,   6.60294){\circle*{0.2}} 
\put(   39.17868,   9.01716){\circle*{0.2}} 
\put(   38.47157,  10.72426){\circle*{0.2}} 
\put(   39.17868,  12.43137){\circle*{0.2}} 
\put(   40.17868,  13.43137){\circle*{0.2}} 
\put(   39.47157,  13.13848){\circle*{0.2}} 
\put(   41.88579,   5.89584){\circle*{0.2}} 
\put(   42.59289,   5.60294){\circle*{0.2}} 
\put(   43.30000,   5.89584){\circle*{0.2}} 
\put(   41.59289,  14.84558){\circle*{0.2}} 
\put(   40.17868,  14.84558){\circle*{0.2}} 
\put(   43.30000,  15.55269){\circle*{0.2}} 
\put(   42.59289,  15.84558){\circle*{0.2}} 
\put(   41.88579,  15.55269){\circle*{0.2}} 
\put(   44.30000,   4.89584){\circle*{0.2}} 
\put(   46.00711,   5.60294){\circle*{0.2}} 
\put(   46.71421,   5.89584){\circle*{0.2}} 
\put(   45.71421,   4.89584){\circle*{0.2}} 
\put(   48.42132,  14.84558){\circle*{0.2}} 
\put(   48.42132,  13.43137){\circle*{0.2}} 
\put(   49.42132,  12.43137){\circle*{0.2}} 
\put(   49.12843,  13.13848){\circle*{0.2}} 
\put(   49.12843,  11.72426){\circle*{0.2}} 
\put(   47.42132,  15.84558){\circle*{0.2}} 
\put(   46.71421,  15.55269){\circle*{0.2}} 
\put(   46.00711,  15.84558){\circle*{0.2}} 
\put(   50.12843,  10.72426){\circle*{0.2}} 
\put(   49.12843,   9.72426){\circle*{0.2}} 
\put(   49.42132,   9.01716){\circle*{0.2}} 
\put(   49.12843,   8.31005){\circle*{0.2}} 
\put(   45.71421,  16.55269){\circle*{0.2}} 
\put(   44.30000,  16.55269){\circle*{0.2}} 
\put(   48.42132,   6.60294){\circle*{0.2}} 
\put(   39.17868,   6.60294){\circle*{0.2}} 
\put(   38.47157,   8.31005){\circle*{0.2}} 
\put(   40.17868,   5.60294){\circle*{0.2}} 
\put(   40.88579,   5.89584){\circle*{0.2}} 
\put(   38.47157,   9.72426){\circle*{0.2}} 
\put(   37.47157,  10.72426){\circle*{0.2}} 
\put(   37.76447,  10.01716){\circle*{0.2}} 
\put(   38.47157,  11.72426){\circle*{0.2}} 
\put(   37.76447,  11.43137){\circle*{0.2}} 
\put(   38.47157,  13.13848){\circle*{0.2}} 
\put(   39.47157,  14.13848){\circle*{0.2}} 
\put(   41.88579,   4.89584){\circle*{0.2}} 
\put(   43.30000,   4.89584){\circle*{0.2}} 
\put(   40.88579,  15.55269){\circle*{0.2}} 
\put(   39.17868,  14.84558){\circle*{0.2}} 
\put(   40.17868,  15.84558){\circle*{0.2}} 
\put(   43.30000,  16.55269){\circle*{0.2}} 
\put(   41.88579,  16.55269){\circle*{0.2}} 
\put(   43.59289,   4.18873){\circle*{0.2}} 
\put(   45.00711,   4.18873){\circle*{0.2}} 
\put(   46.71421,   4.89584){\circle*{0.2}} 
\put(   47.71421,   5.89584){\circle*{0.2}} 
\put(   49.42132,  14.84558){\circle*{0.2}} 
\put(   49.12843,  15.55269){\circle*{0.2}} 
\put(   48.42132,  15.84558){\circle*{0.2}} 
\put(   49.12843,  14.13848){\circle*{0.2}} 
\put(   50.12843,  13.13848){\circle*{0.2}} 
\put(   50.12843,  11.72426){\circle*{0.2}} 
\put(   46.71421,  16.55269){\circle*{0.2}} 
\put(   50.83553,  11.43137){\circle*{0.2}} 
\put(   50.12843,   9.72426){\circle*{0.2}} 
\put(   50.83553,  10.01716){\circle*{0.2}} 
\put(   50.12843,   8.31005){\circle*{0.2}} 
\put(   49.12843,   7.31005){\circle*{0.2}} 
\put(   45.00711,  17.25980){\circle*{0.2}} 
\put(   44.30000,  17.55269){\circle*{0.2}} 
\put(   43.59289,  17.25980){\circle*{0.2}} 
\put(   49.42132,   6.60294){\circle*{0.2}} 
\put(   48.42132,   5.60294){\circle*{0.2}} 
\put(   49.12843,   5.89584){\circle*{0.2}} 
\put(   38.47157,   5.89584){\circle*{0.2}} 
\put(   39.17868,   5.60294){\circle*{0.2}} 
\put(   38.47157,   7.31005){\circle*{0.2}} 
\put(   37.47157,   8.31005){\circle*{0.2}} 
\put(   37.76447,   9.01716){\circle*{0.2}} 
\put(   39.47157,   4.89584){\circle*{0.2}} 
\put(   40.88579,   4.89584){\circle*{0.2}} 
\put(   36.76447,  10.01716){\circle*{0.2}} 
\put(   36.76447,  11.43137){\circle*{0.2}} 
\put(   37.76447,  12.43137){\circle*{0.2}} 
\put(   37.47157,  13.13848){\circle*{0.2}} 
\put(   38.47157,  14.13848){\circle*{0.2}} 
\put(   42.59289,   4.18873){\circle*{0.2}} 
\put(   40.88579,  16.55269){\circle*{0.2}} 
\put(   39.17868,  15.84558){\circle*{0.2}} 
\put(   38.47157,  15.55269){\circle*{0.2}} 
\put(   39.47157,  16.55269){\circle*{0.2}} 
\put(   42.59289,  17.25980){\circle*{0.2}} 
\put(   43.59289,   3.18873){\circle*{0.2}} 
\put(   44.30000,   3.48162){\circle*{0.2}} 
\put(   46.00711,   4.18873){\circle*{0.2}} 
\put(   45.00711,   3.18873){\circle*{0.2}} 
\put(   47.71421,   4.89584){\circle*{0.2}} 
\put(   47.42132,   4.18873){\circle*{0.2}} 
\put(   50.12843,  15.55269){\circle*{0.2}} 
\put(   50.12843,  14.13848){\circle*{0.2}} 
\put(   49.12843,  16.55269){\circle*{0.2}} 
\put(   47.71421,  16.55269){\circle*{0.2}} 
\put(   50.83553,  12.43137){\circle*{0.2}} 
\put(   47.42132,  17.25980){\circle*{0.2}} 
\put(   46.71421,  17.55269){\circle*{0.2}} 
\put(   46.00711,  17.25980){\circle*{0.2}} 
\put(   51.83553,  11.43137){\circle*{0.2}} 
\put(   51.54264,  10.72426){\circle*{0.2}} 
\put(   50.83553,   9.01716){\circle*{0.2}} 
\put(   51.83553,  10.01716){\circle*{0.2}} 
\put(   50.12843,   7.31005){\circle*{0.2}} 
\put(   50.83553,   7.60294){\circle*{0.2}} 
\put(   45.00711,  18.25980){\circle*{0.2}} 
\put(   43.59289,  18.25980){\circle*{0.2}} 
\put(   50.12843,   5.89584){\circle*{0.2}} 
\put(   49.12843,   4.89584){\circle*{0.2}} 
\put(   37.47157,   5.89584){\circle*{0.2}} 
\put(   38.47157,   4.89584){\circle*{0.2}} 
\put(   37.76447,   6.60294){\circle*{0.2}} 
\put(   37.47157,   7.31005){\circle*{0.2}} 
\put(   36.76447,   7.60294){\circle*{0.2}} 
\put(   36.76447,   9.01716){\circle*{0.2}} 
\put(   40.17868,   4.18873){\circle*{0.2}} 
\put(   41.59289,   4.18873){\circle*{0.2}} 
\put(   36.05736,  10.72426){\circle*{0.2}} 
\put(   36.76447,  12.43137){\circle*{0.2}} 
\put(   37.47157,  14.13848){\circle*{0.2}} 
\put(   37.76447,  14.84558){\circle*{0.2}} 
\put(   42.59289,   3.18873){\circle*{0.2}} 
\put(   41.59289,  17.25980){\circle*{0.2}} 
\put(   40.88579,  17.55269){\circle*{0.2}} 
\put(   40.17868,  17.25980){\circle*{0.2}} 
\put(   38.47157,  16.55269){\circle*{0.2}} 
\put(   42.59289,  18.25980){\circle*{0.2}} 
\put(   44.30000,   2.48162){\circle*{0.2}} 
\put(   46.00711,   3.18873){\circle*{0.2}} 
\put(   46.71421,   3.48162){\circle*{0.2}} 
\put(   45.71421,   2.48162){\circle*{0.2}} 
\put(   48.42132,   4.18873){\circle*{0.2}} 
\put(   47.42132,   3.18873){\circle*{0.2}} 
\put(   50.12843,  16.55269){\circle*{0.2}} 
\put(   50.83553,  14.84558){\circle*{0.2}} 
\put(   50.83553,  13.43137){\circle*{0.2}} 
\put(   49.12843,  17.55269){\circle*{0.2}} 
\put(   48.42132,  17.25980){\circle*{0.2}} 
\put(   51.83553,  12.43137){\circle*{0.2}} 
\put(   51.54264,  13.13848){\circle*{0.2}} 
\put(   47.42132,  18.25980){\circle*{0.2}} 
\put(   46.00711,  18.25980){\circle*{0.2}} 
\put(   52.54264,  10.72426){\circle*{0.2}} 
\put(   51.83553,   9.01716){\circle*{0.2}} 
\put(   51.54264,   8.31005){\circle*{0.2}} 
\put(   50.83553,   6.60294){\circle*{0.2}} 
\put(   45.71421,  18.96690){\circle*{0.2}} 
\put(   44.30000,  18.96690){\circle*{0.2}} 
\put(   50.12843,   4.89584){\circle*{0.2}} 
\put(   37.47157,   4.89584){\circle*{0.2}} 
\put(   36.76447,   6.60294){\circle*{0.2}} 
\put(   37.76447,   4.18873){\circle*{0.2}} 
\put(   39.17868,   4.18873){\circle*{0.2}} 
\put(   36.05736,   8.31005){\circle*{0.2}} 
\put(   35.76447,   9.01716){\circle*{0.2}} 
\put(   36.05736,   9.72426){\circle*{0.2}} 
\put(   40.17868,   3.18873){\circle*{0.2}} 
\put(   40.88579,   3.48162){\circle*{0.2}} 
\put(   41.59289,   3.18873){\circle*{0.2}} 
\put(   35.05736,  10.72426){\circle*{0.2}} 
\put(   36.05736,  11.72426){\circle*{0.2}} 
\put(   35.76447,  12.43137){\circle*{0.2}} 
\put(   36.76447,  13.43137){\circle*{0.2}} 
\put(   36.05736,  13.13848){\circle*{0.2}} 
\put(   36.76447,  14.84558){\circle*{0.2}} 
\put(   37.76447,  15.84558){\circle*{0.2}} 
\put(   41.88579,   2.48162){\circle*{0.2}} 
\put(   43.30000,   2.48162){\circle*{0.2}} 
\put(   41.59289,  18.25980){\circle*{0.2}} 
\put(   40.17868,  18.25980){\circle*{0.2}} 
\put(   39.17868,  17.25980){\circle*{0.2}} 
\put(   37.47157,  16.55269){\circle*{0.2}} 
\put(   38.47157,  17.55269){\circle*{0.2}} 
\put(   37.76447,  17.25980){\circle*{0.2}} 
\put(   43.30000,  18.96690){\circle*{0.2}} 
\put(   41.88579,  18.96690){\circle*{0.2}} 
\put(   44.30000,   1.48162){\circle*{0.2}} 
\put(   45.00711,   1.77452){\circle*{0.2}} 
\put(   46.71421,   2.48162){\circle*{0.2}} 
\put(   49.42132,   4.18873){\circle*{0.2}} 
\put(   48.42132,   3.18873){\circle*{0.2}} 
\put(   49.12843,   3.48162){\circle*{0.2}} 
\put(   50.83553,  17.25980){\circle*{0.2}} 
\put(   50.12843,  17.55269){\circle*{0.2}} 
\put(   50.83553,  15.84558){\circle*{0.2}} 
\put(   51.83553,  14.84558){\circle*{0.2}} 
\put(   51.54264,  14.13848){\circle*{0.2}} 
\put(   48.42132,  18.25980){\circle*{0.2}} 
\put(   52.54264,  13.13848){\circle*{0.2}} 
\put(   52.54264,  11.72426){\circle*{0.2}} 
\put(   46.71421,  18.96690){\circle*{0.2}} 
\put(   53.54264,  10.72426){\circle*{0.2}} 
\put(   53.24975,  11.43137){\circle*{0.2}} 
\put(   52.54264,   9.72426){\circle*{0.2}} 
\put(   53.24975,  10.01716){\circle*{0.2}} 
\put(   52.54264,   8.31005){\circle*{0.2}} 
\put(   51.54264,   7.31005){\circle*{0.2}} 
\put(   51.83553,   6.60294){\circle*{0.2}} 
\put(   50.83553,   5.60294){\circle*{0.2}} 
\put(   51.54264,   5.89584){\circle*{0.2}} 
\put(   45.00711,  19.67401){\circle*{0.2}} 
\put(   44.30000,  19.96690){\circle*{0.2}} 
\put(   50.83553,   4.18873){\circle*{0.2}} 
\put(   36.76447,   4.18873){\circle*{0.2}} 
\put(   36.76447,   5.60294){\circle*{0.2}} 
\put(   35.76447,   6.60294){\circle*{0.2}} 
\put(   36.05736,   5.89584){\circle*{0.2}} 
\put(   36.05736,   7.31005){\circle*{0.2}} 
\put(   37.76447,   3.18873){\circle*{0.2}} 
\put(   38.47157,   3.48162){\circle*{0.2}} 
\put(   39.17868,   3.18873){\circle*{0.2}} 
\put(   35.05736,   8.31005){\circle*{0.2}} 
\put(   35.05736,   9.72426){\circle*{0.2}} 
\put(   39.47157,   2.48162){\circle*{0.2}} 
\put(   40.88579,   2.48162){\circle*{0.2}} 
\put(   34.35025,  10.01716){\circle*{0.2}} 
\put(   35.05736,  11.72426){\circle*{0.2}} 
\put(   34.35025,  11.43137){\circle*{0.2}} 
\put(   35.05736,  13.13848){\circle*{0.2}} 
\put(   36.05736,  14.13848){\circle*{0.2}} 
\put(   35.76447,  14.84558){\circle*{0.2}} 
\put(   36.76447,  15.84558){\circle*{0.2}} 
\put(   36.05736,  15.55269){\circle*{0.2}} 
\put(   41.88579,   1.48162){\circle*{0.2}} 
\put(   42.59289,   1.77452){\circle*{0.2}} 
\put(   43.30000,   1.48162){\circle*{0.2}} 
\put(   40.88579,  18.96690){\circle*{0.2}} 
\put(   39.17868,  18.25980){\circle*{0.2}} 
\put(   39.47157,  18.96690){\circle*{0.2}} 
\put(   36.76447,  17.25980){\circle*{0.2}} 
\put(   37.76447,  18.25980){\circle*{0.2}} 
\put(   43.30000,  19.96690){\circle*{0.2}} 
\put(   42.59289,  19.67401){\circle*{0.2}} 
\put(   41.88579,  19.96690){\circle*{0.2}} 
\put(   43.59289,    .77452){\circle*{0.2}} 
\put(   45.00711,    .77452){\circle*{0.2}} 
\put(   29.47619,  10.00000){\circle*{0.2}} 
\put(   27.47619,  10.00000){\circle*{0.2}} 
\put(   29.18329,  10.70711){\circle*{0.2}} 
\put(   27.76908,   9.29289){\circle*{0.2}} 
\put(   28.47619,  11.00000){\circle*{0.2}} 
\put(   28.47619,   9.00000){\circle*{0.2}} 
\put(   27.76908,  10.70711){\circle*{0.2}} 
\put(   29.18329,   9.29289){\circle*{0.2}} 
\put(   28.47619,  10.00000){\circle*{0.2}} 
\end{picture}		   
\caption{{\it Left:} The strip ${\bf S}$ and the window ${\bf W}$ in the case of a
1D physical space ${\bf E}$ embedded into a 3D superspace.
{\it Centre:} A one-shell $D_8$-cluster $\mathcal{C}$. 
{\it Right:} A fragment of the set $\mathcal{Q}$ defined by $\mathcal{C}$.
The nearest neighbours of any point $q$ of $\mathcal{Q}$ belong to $q+\mathcal{C}$, 
which is a copy of $\mathcal{C}$ with the center at point $q$. 
Therefore, $\mathcal{C}$ is a covering cluster for $\mathcal{Q}$.} 
\end{figure}

In the case $n=2$, the vector defined by the formal determinant
\begin{equation} y=\left| \begin{array}{ccccccc}
e_1 & e_2 & e_3 & e_4 & e_5 & ... & e_k\\
0 & 0 & 0 & 1 & 0 & ... & 0\\ 
0 & 0 & 0 & 0 & 1 & ... & 0\\ 
... & ... & ...& ...& ...& ...\\ 
0 & 0 & 0 & 0 & 0 &  ...& 1\\
v_{11} & v_{12} & v_{13} & v_{14} & v_{15} & ... & v_{1k}\\
v_{21} & v_{22} & v_{23} & v_{24} & v_{25} & ... & v_{2k}\\
\end{array} \right| 
= \left| \begin{array}{ccc}
e_1 & e_2 & e_3\\
v_{11} & v_{12} & v_{13}\\
v_{21} & v_{22} & v_{23}
\end{array} \right| \end{equation}
is a vector orthogonal to the vectors $e_4$, $e_5$, ...,  $e_k$, $w_1$, $w_2$, and
\begin{equation} \langle x,y\rangle =
\left| \begin{array}{ccc}
x_1& x_2 & x_3\\
v_{11} & v_{12} & v_{13}\\
v_{21} & v_{22} & v_{23}
\end{array} \right| \end{equation}
for any $x\in \mathbb{R}^k$. The vector  $y$ belongs to ${\bf E}^\perp $, and since
$e_i-\pi ^\perp e_i$ is a linear 
combination of $w_1$ and $w_2$, it is also orthogonal to $\pi ^\perp e_4$, 
$\pi ^\perp e_5$, ..., $\pi ^\perp e_k$. Therefore, $y$ is orthogonal 
to the $(k-3)$-faces of ${\bf W}$ labelled by $(1,2,3)$. Similar results can be 
obtained for any $(i_1,i_2,i_3)\in \mathcal{I}$.

Consider the {\it strip} ${\bf S}$ corresponding to the window ${\bf W}$ (see figure 1)
\begin{equation} 
{\bf S}=\{ x\in \mathbb{R}^k\ |\ \pi ^\perp x\in {\bf W} \ \} 
\end{equation}
and define for each $(i_1,i_2,...,i_{n+1})\in \mathcal{I}$ the number
\begin{equation} 
d_{i_1i_2...i_{n+1}}=\max_{\alpha _j \in \{ -1/2,\, 1/2\}}
\left| 
\begin{array}{cccc}
\alpha _1& \alpha _2 & ...&\alpha _{n+1} \\
v_{1i_1} & v_{1i_2} & ...& v_{1i_{n+1}}\\
v_{2i_1} & v_{2i_2} & ... & v_{2i_{n+1}}\\
... & ... & ... & ...\\
v_{ni_1} & v_{ni_2} & ... & v_{ni_{n+1}}
\end{array} \right|.
\end{equation}
A point $x=(x_1,x_2,...,x_k)\in \mathbb{R}^k$ 
belongs to the strip ${\bf S}$ if and only if 
\begin{equation} -d_{i_1i_2...i_{n+1}}\leq 
\left| %
		   
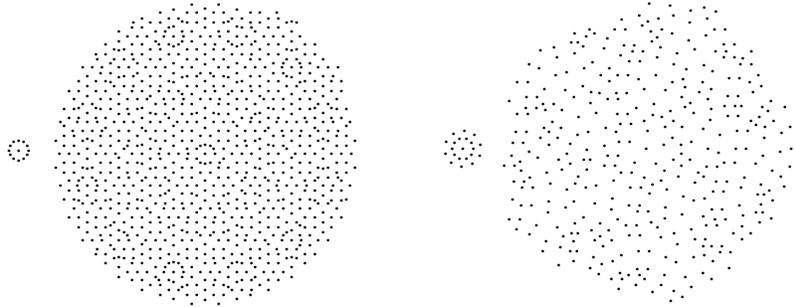
\caption{{\it Left:} A one-shell $D_{12}$-cluster and a fragment of the corresponding 
quasiperiodic set. 
{\it Right:} A fragment of the quasiperiodic set defined by a two-shell $D_{10}$-cluster,
obtained by using strip projection method in a ten-dimensional superspace. 
The starting cluster is a covering cluster, 
but for most of the points the occupation is extremely low.} 
\end{figure}

The set defined in terms of the strip projection method 
\begin{equation} \mathcal{Q}=\mathcal{P}({\bf S}\cap \mathbb{Z}^k)=
\{ \mathcal{P}x\ |\ \ x\in {\bf S}\cap \mathbb{Z}^k\ \} \end{equation}
by using the strip ${\bf S}$ defined above and the  
hyper-lattice $\mathbb{Z}^k$ is a discrete set. 
If $G$ is one of the groups occurring in quasicrystal physics
then $\mathcal{Q}$ is a quasiperiodic set, and has all the properties 
of the sets obtained by projection \cite{md,kp}. 
We have 
\begin{equation} 
\mathcal{P}e_i=(\langle e_i,w_1\rangle , \langle e_i, w_2\rangle ,
...,\langle e_i,w_n\rangle )
=(v_{1i},v_{2i},...,v_{ni})=v_i 
\end{equation}
for any $i\in \{ 1,2,...,k\}$, whence
\begin{equation} \begin{array}{l}
 \mathcal{P}(\{x\pm e_1,\, x\pm e_2,\, ...,\, x\pm e_k \}\cap {\bf S})\\[1mm]
 \mbox{}\ \ \ \subseteq \{ \mathcal{P}x\pm v_1,\, \mathcal{P}x\pm v_2\, ...,\, 
\mathcal{P}x\pm v_k\}=\mathcal{P}x+\mathcal{C} 
\end{array} \end{equation}
that is, the `arithmetical' neighbours \cite{md} of any point
$\mathcal{P}x\in \mathcal{Q}$ belong to the translated copy 
$\mathcal{P}x+\mathcal{C}$ of $\mathcal{C}$. Therefore, the starting cluster
$\mathcal{C}$ can be regarded as a {\it covering cluster} 
(\cite{kp}, page 16) of the point set $\mathcal{Q}$. Some examples are shown 
in figure 1 and  figure 2. A presentation of our algorithm in the
particular case of a two-shell $D_{10}$-cluster can be found in \cite{cp1}.\\[5mm]
\noindent{\bf 4. Concluding remarks}\\[3mm]
It is known that, generally, the discrete sets $\mathcal{Q}$ considered in the 
previous section can also be defined as multi-component model sets \cite{bm,nc1} 
by using superspaces of smaller dimension and root lattices, but generally we 
have to use a large number of very complicated windows \cite{nc1}.

The starting cluster $\mathcal{C}$ is a covering cluster for the set $\mathcal{Q}$,
but unfortunately, for most of the points of $\mathcal{Q}$ the occupation is
extremely low. Therefore, our discrete quasiperiodic sets can not be used directly
in the description of atomic positions in quasicrystals.\\[3mm]
{\bf Acknowledgment.} This work was supported by CERES project 4-129.

\end{document}